# Effect of Boron substitution on the superconductivity of non-oxide perovskite MgCNi$_3$

Anuj Kumar[1,2], Rajveer Jha[1], R. P. Tandon[2] and V. P. S. Awana[1,*]

[1]*Quantum Phenomena and Application Division, National Physical Laboratory (CSIR) Dr. K. S. Krishnan Road, New Delhi-110012, India*

[2]*Department of Physics and Astrophysics, University of Delhi, North Campus New Delhi-110007, India*

## Abstract

We report synthesis, structural and magnetic (*DC* and *AC*) properties of Boron substituted MgCNi$_3$ superconductor. A series of polycrystalline bulk samples Mg$_{1.2}$C$_{1.6-x}$B$_x$Ni$_3$ (x = 0.0, 0.08 and 0.16) is synthesized through standard solid-state reaction route, which are found to crystallize in cubic perovskite structure with space group *Pm3m*. Rietveld analysis of observed XRD data show that lattice parameters expand from *a* = 3.8106 (4) Å for pure, to 3.8164 (2) Å and 3.8173 (5) Å for 5% and 10% Boron substituted samples respectively. *DC* magnetization exhibited superconducting transition ($T_c$) at around 7.3 K for pure sample, and the same decreases slightly with Boron substitution. The lower critical field ($H_{c1}$) at 2 K is around 150 Oe for pure sample, which increases slightly with Boron substitution. For pure sample the upper critical field ($H_{c2}$) being determined from *AC* susceptibility measurements is 11.6 kOe and 91.70 kOe with 50% and 90% diamagnetism criteria respectively, which decreases to 5.57 kOe and 42.5 kOe respectively for 10% Boron substituted sample. 10% Boron substitution at Carbon site has decreased both the $H_{c2}$ and $T_c$. On the other hand lower critical field ($H_{c1}$) at 2 K is slightly increased from around 150 Oe for pure sample, to 200 Oe for 10% Boron substituted sample. Seemingly, the Carbon site Boron substitution induced disorder though has increased slightly the $H_{c1}$ but with simultaneous decrease in superconducting transition temperature ($T_c$) and upper critical field ($H_{c2}$). The high relative proportion of Ni in studied MgCNi$_3$ suggests that magnetic interactions are important and non-oxide perovskite structure make it interesting.

# I. Introduction

The discovery of superconductivity in the $MgB_2$ compound with critical temperature $T_c \sim$ 39 K [1] renewed the interest for the search of new inter-metallic compounds with high superconducting transition temperatures. In 2001, $MgB_2$ was the only superconducting compound other than cuprates having the maximum critical temperature $T_c$. Later on, another inter-metallic compound containing Mg metal was discovered named as $MgCNi_3$ with superconducting transition temperature $T_c \sim 8$ K [2]. $MgCNi_3$ crystallizes in a classical cubic perovskite structure (space group $Pm3m$), which is related to the structure of high $T_c$ cuprate superconductors. Perovskite inter-metalics with formula $AXM_3$, are related to both classical inter-metallics $AuBCu_3$ and oxide peroveskite such as $CaTiO_3$. The interstitial X elements, typically B, C, or N can be considered either as entering the body centered position of the $AuBCu_3$ cell, or as being in the transition metal position in the perovskite cell. The structural and electronic analogies between $MgCNi_3$ and $CaTiO_3$ are more extensive than might be initially perceived. $MgCNi_3$ is oxygen free perovskite superconducting compound. The superconducting transition temperature $T_c$ of $MgCNi_3$ is low (around 8 K) compared to high $T_c$ cuprates like $La_{1.85}Sr_{0.15}CuO_4$ ($T_c = 35$ K) [3] and $YBa_2Cu_3O_{7-\delta}$ ($T_c = 92$ K) [4] but in to those of related inter-metallic superconductors, including $LuNi_2B_2C$ ($T_c = 16$ K) [5, 6] and $YPd_2B_2C$ ($T_c = 23$ K) [7]. The observance of superconductivity in Ni based compound is not unprecedented and already reported in $LaNiC_2$ [8] and binary alloy of Bi and Ni ($Bi_3Ni$) [9] etc. are some known Ni based superconductors. There are some evidences suggesting that $MgCNi_3$ may represent a new type of unconventional superconductivity and many fundamental questions about the nature of its superconducting state still remain unanswered. Superconductivity and ferromagnetism are believed to be incompatible over any temperature range until, except a few specific examples such as $UGe_2$, $ZnZr_2$ and $RuSr_2GdCu_2O_{8-\delta}$. $MgCNi_3$ is believed to be at the border line case between superconductivity and ferromagnetism [10]. This system provides opportunity to probe coupling and possible co-existence of superconductivity in vicinity of ferromagnetism. The perovskite $MgCNi_3$ is special in that it is neither an oxide nor does it contain copper. Full occupancy of divalent Ni metal in $MgCNi_3$ suggests that magnetic interaction may contribute to ferromagnetism rather than the superconductivity. Band-structure calculations reveal a narrow peak in the vicinity of the Fermi energy ($E_F$) [11-14]. The superconducting properties of the $MgCNi_3$ are associated with the occurrence of an intense peak in the density of Ni $3d$ states at the Fermi level. Calculations suggest that introducing electrons or hole dopants into $MgCNi_3$ will result in a decrease or increase in N ($E_F$), respectively. In the former case Ni substituting by Cu [13, 14], one can expect that superconductivity of the system deteriorate. While, with holes (Ni by Fe) will increase the N ($E_F$). A large N ($E_F$) is not always good for superconductivity, because the same may introduce spin fluctuation or magnetic order in the system. Shim et. al., [14] showed that the Fermi surface is composed of two bands and also proposed the strong narrow density of states (DOS) peak, located just below the $E_F$, which is corresponding to the anti-bonding state of Ni $3d$ and C $2p$ state. Therefore, hole doping of $MgCNi_3$ to move the Fermi level into the DOS peak should be of interest. Well, neither increase in superconducting transition temperature $T_c$ nor the ferromagnetism is observed with the partial substitution of Co, Fe, Ru, or Mn for Ni site in $MgCNi_3$ [15-17]. But Carbon atom plays an important role in $MgCNi_3$ and affects its superconductivity [18]. Only one report in literature, deals with Boron ($^{11}$B) substitution at C site in $MgCNi_3$ [19]. The synthesis of $MgCNi_3$ requires both Mg and C in excess amount due to the volatile nature of Mg and to ensure C incorporation [2, 18, 20]. So the controlled doping of both the Mg and the C site is difficult and crystal structure analysis (Rietveld analysis) is required to determine the true composition. The Carbon site in $MgCNi_3$ perovskite was found underoccupied as reported earlier [21, 22]. The optimization of the process for pure $MgCNi_3$ end compound was reported by us very recently [23]. There are only two possible candidates for substitution at Carbon site i.e., Boron and Nitrogen. The present study deals with the impact of Boron substitution at Carbon site in $MgCNi_3$. Though the superconducting transition ($T_c$) and upper critical field ($H_{c2}$) are decreased, the lower critical field ($H_{c1}$) is slightly improved with B substitution at C site in $MgCNi_3$. Here we just report a study of superconducting parameters like transition temperature ($T_c$), lower critical field ($H_{c1}$) and upper critical field ($H_{c2}$) of $MgC_{1-x}B_xNi_3$ (x = 0%, 5% and 10%).

## II. Experimental Details

A series of bulk polycrystalline samples with compositions $Mg_{1.2}C_{1.6-x}B_xNi_3$ (x = 0.0, 0.08 and 0.16) was synthesized through standard solid state reaction route. Excess Mg and C are needed in order to get the optimal Mg and C content for stoichiometric $MgCNi_3$ [2, 16]. The initial elements were pure Mg flakes (99% Riedal-de Haen), Ni (99.99% Sigma Aldrich), amorphous Carbon powder (Alfa Aesar) and Boron metal (99.9% Alfa Aesar) powder. The stoichiometric amounts of high purity ingredients (Mg, amorphous C, Boron and Ni) are ground thoroughly in an agate mortar and pelletized using hydraulic press. These pressed bar shaped pallets are put in an iron alloy (Fe+Mn+Cr+C, melting point 1400°C-1500°C) capsule and then initially heated at 600°C for 2 h at the rate of 10°C/minute in continuous flow of Argon gas atmosphere of purity (99.9%), followed by heating at higher temperature 950°C for 3 h and finally cooled down to room temperature in the same atmosphere. Both ends of iron alloy capsule are kept open for passage of the pure Ar gas. This iron alloy tube is used to control the evaporation of Mg and C, during heat treatment. The resultant sample is in powder form. We did this earlier in case of synthesis of $MgB_2$ as well and obtained high purity compound [24]. X-ray diffraction was performed at room temperature in the scattering angular (2θ) range of 20°-80° in equal steps of 0.02° using Rigaku Diffractometer with Cu $K_α$ (λ = 1.54 Å) radiation. Rietveld analysis was performed using the standard FullProf program. All magnetic (*DC*, *AC* susceptibility and isothermal magnetization) measurements are carried out using ACMS assembly on physical property measurement system (PPMS-14 Tesla, Quantum Design-USA) down to 2 K with varying fields of up to 140 kOe.

## III. Results and Discussion

Figure 1 shows the observed and fitted X-ray diffraction pattern for pure, 5% and 10% substituted $Mg_{1.2}C_{1.6-x}B_xNi_3$ compound. The structural analysis was performed by Rietveld refinement by using the *FullProf* Program. The Rietveld analysis confirmed single phase formation of all samples in cubic oxygen free perovskite strcuture with the same space group *Pm3m*. The site occupancy and coordinates position for the atoms are observed as Mg: 1*a* (0, 0, 0), C: 1*b* (0.5, 0.5, 0.5) and Ni 3*c* (0, 0.5, 0.5). Un-reacted Ni impurity is not seen within the resolution limit of XRD, except a minute impurity for 10% Boron substituted sample at around 44.5 degree. After Rietveld analysis of all samples, an expansion of the lattice parameter with the increase of systematic Boron content is observed. The small expansion in lattice is due to the substitution of little bit larger atomic radii of Boron atom (85 pm) at the Carbon atom (77.2 pm) site. All Rietveld refined parameters like lattice parameter, Wyckoff position and refined occupancies corresponding to each atom are shown in Tables I, II and III. The site occupancy corresponding to each refined atoms/ions is not fixed but is left free for the best fit of observed and calculated patterns. The site occupancy provided in Table I, II and III are generated from best fits of XRD patterns in *FULLPROF* program. If the site occupancy factor is exactly 1 (sof = 1) it means that, every equivalent position for x, y, and z is occupied by that atom. On the other hand if the site occupancy is greater than 1 (sof > 1), means some of the sites are vacant.

The temperature dependent *DC* magnetization of phase pure samples $Mg_{1.2}C_{1.6-x}B_xNi_3$ (x = 0.0, 0.08 and 0.16) is shown in Figure 2. The sample was cooled in zero magnetic fields down to lowest accessible temperature (2 K). After stabilization of temperature at 2 K, a *DC* magnetic field of 10 Oe is applied and the magnetization is recorded as a function of temperature (called *ZFC* mode) up to 8 K. Further, the measurement continued while the temperature was again decreased back to 2 K (*FC* mode), keeping the same *DC* magnetic field. Pristine sample shows the bulk superconductivity with an onset temperature of ($T_c^{onset}$) of 7.3 K. It is observed that Boron substitution at Carbon site decreases the superconducting transition temperature ($T_c$) from 7.3 K (x =0.0) to 6.6 K (x= 0.16). The numerical values of superconducting transition $T_c$ for all the studied samples are given in Table IV. This change in $T_c$ is relatively small compare to that in $Mg(B_{1-x}C_x)_2$, in which 10% Carbon substitution decreases $T_c$ from 39 K to around 33 K [24]. This small suppression of superconducting transition can be explained on the basis of band structure calculations. For the parent compound named $MgCNi_3$ the Fermi energy $E_F$ is of the high-energy side with an intense peak of the density of Ni *3d* states [11-14, 25]. Doping the compound with Boron (hole doping) reduces the number of valence electrons and shifts the Fermi level to the lower energy side, and thus increasing the DOS at the Fermi level N ($E_F$). According to the BCS Theory, $T_c$ may rise with the increase in N ($E_F$). However, an increase in the DOS leads to enhanced Stoner exchange parameter S = N ($E_F$)$I_{ex}$ ($I_{ex}$ is the exchange integral) due to which induced spin fluctuation or magnetic order in the system destroys the superconducting pairing state [11, 25]. Details study of Density of states (DOS) calculations using density functional theory (DFT) with and without spin fluctuations of $MgCNi_3$ is shown in ref. 23.

The low field (up to 1000 Oe) isothermal magnetization *MH* plots of studied Boron substituted samples $Mg_{1.2}C_{1.6-x}B_xNi_3$ (x= 0.0, 0.08 and 0.16) at temperature *T* = 2, 4 and 6 K are shown in Figure 3(a), (b) and (c) respectively. To determine the lower critical field ($H_{c1}$), we draw a tangent line from the origin on the curve. Lower critical field $H_{c1}$ is defined as deviation point between the tangent line and the curve at 2 K (shown in figure 3(a), (b) and (c)). Same method is employed to determine the $H_{c1}$ at 4 and 6 K. Thus, determined lower critical field $H_{c1}$ for pure sample is around 150 Oe at 2 K. The $H_{c1}$ decreases monotonically with increase in temperature to 85 and 35 Oe at 4 and 6 K respectively. Same isothermal magnetization *MH* behavior is followed by Boron substituted samples. The $H_{c1}$ for 5% Boron substituted sample is ~160 Oe at 2 K and the same decreases monotonically with increase in temperature to 95 and 50 Oe at 4 and 6 K respectively. For the 10% Boron substituted sample the observed $H_{c1}$ is 200 Oe at 2 K and the same is decreased with increase in temperature to 160 Oe and 70 Oe at 4 and 6 K respectively. The numerical values of $H_{c1}$ for all the studied samples are given in Table IV. Hence, it is clear from the observed data the lower critical field ($H_{c1}$) increases slightly with the 5% and 10% substitution of Boron concentration at Carbon site in $MgCNi_3$ matrix which is possibly due to intrinsic disorder in the system. Though, the exact mechanism of slightly increasing the lower critical field $H_{c1}$ with Boron substitution at Carbon site in $MgCNi_3$ may be unknown and yet debatable.

Figure 4(a), (b) and (c) shows the upper critical field ($H_{c2}$) being determined from in-field *AC* susceptibility measurements for pristine, 5% and 10% Boron substituted $MgCNi_3$ with applied frequency 333 *Hz* and *AC* driven field amplitude of 10 Oe. All samples are measured up to the 140 kOe external *DC* magnetic field at temperature 2 K. The upper critical field ($H_{c2}$) for pure sample is 11.6 kOe and 91.70 kOe with 50% and 90% diamagnetism criteria respectively. On the other hand the upper critical field ($H_{c2}$) for 5% and 10% Boron substituted samples is 9.43 kOe (50% diamagnetism), 58.44 kOe (90% diamagnetism) and 5.57 kOe (50% diamagnetism), 42.5 kOe (90% diamagnetism) respectively. The numerical values of $H_{c2}$ at 2 K for all studied samples are given in Table IV. The values of $H_{c2}$ as observed from the *AC* susceptibility measurements are below the paramagnetic limit $H_P = 1.84T_c$, i.e., around 130 kOe, suggesting that the Zeeman pair breaking mechanism is not effective in this case. Decrease of the upper critical field on Boron substitution can be explained in terms of the $H_{c2}$ dependence on the effective coherence length $\xi$ as $H_{c2} \propto \Phi_0/\xi^2$ where $\Phi_0$ is the flux quantum. For a dirty superconductors $1/\xi = 1/\xi_0 + 1/l$, where $\xi_0$ is Pippard coherence length and $l$ is the mean free path. On the other hand $\xi_0$ also related to the critical temperature as $T_c = a\hbar v_0/k\xi_0$, where 'a' is constant, $\hbar = h/2\pi$, where $h$ is a plank constant, $v_0$ is the velocity at Fermi level and $k$ is the Boltzmann's constant. Substitution leads to the shortening of the mean free path [26], but simultaneously for higher substitution, starts to expend $\xi_0$, which leads to the decrease of upper

critical field. It is clear that the upper critical field ($H_{c2}$) decreases drastically with the Boron substitution. It seems that Carbon site Boron substituted disorder, which has been very effective in reverse case of Boron site Carbon substitution in $MgB_2$ superconductor [24], is not helping in case of $MgCNi_3$. On the other hand lower critical field ($H_{c1}$) is slightly improved for Boron substituted $MgCNi_3$.

## IV. Conclusion

Polycrystalline $Mg_{1.2}C_{1.6-x}B_xNi_3$ (x = 0.0, 0.08 and 0.16) samples were synthesized and found to crystallize in cubic structure with space group *Pm3m*. In summary, we have measured superconducting transition ($T_c$), Lower critical field ($H_{c1}$) and Upper critical field ($H_{c2}$) for a series of Boron substituted $MgCNi_3$ compound. The results can be summarized as follows (a) C site B substitution in $MgCNi_3$ is successful till 10%, (b) Superconducting transition temperature ($T_c$) decreases with increase in B content at C site and (c) though the lower critical field ($H_{c1}$) is slightly increased, while the upper critical field ($H_{c2}$) is decreased. Measured physical property parameters (superconducting transition $T_c$, lower critical field $H_{c1}$ and upper critical field $H_{c2}$) are shown in a table IV.

## Acknowledgements

The authors would like to thank Prof. R. C. Budhani (Director, NPL) for his constant support and encouragement during the work. One of the authors Mr. Anuj Kumar would like to thanks Council of Scientific and Industrial Research (CSIR), for providing financial support through Senior Research Fellowship (SRF) to pursue his Ph.D. This work is also financially supported by Department of Science and Technology (DST-SERC) New Delhi, India.

**Table I** $Mg_{1.2}C_{1.6}Ni_3$: Lattice parameters and Unit cell volume: $a = 3.8106$ (3) Å, Vol. 55.333 Å$^3$ and $\chi^2 = 3.53$

| Atom | Site | x | y | z | Site Occupancy |
|------|------|--------|--------|--------|----------------|
| Mg   | 1a   | 0.0000 | 0.0000 | 0.0000 | 1.0641 |
| C    | 1b   | 0.5000 | 0.5000 | 0.5000 | 0.9843 |
| Ni   | 3c   | 0.0000 | 0.5000 | 0.5000 | 1.0244 |

**Table II** $Mg_{1.2}C_{1.52}B_{0.08}Ni_3$: Lattice parameters and Unit cell volume: $a = 3.8164$ (5) Å, Vol. 55.586 Å$^3$ and $\chi^2 = 4.26$

| Atom | Site | x | y | z | Site Occupancy |
|------|------|--------|--------|--------|----------------|
| Mg   | 1a   | 0.0000 | 0.0000 | 0.0000 | 1.0327 |
| C    | 1b   | 0.5000 | 0.5000 | 0.5000 | 0.9831 |
| Ni   | 3c   | 0.0000 | 0.5000 | 0.5000 | 1.0134 |
| B    | 1b   | 0.5000 | 0.5000 | 0.5000 | 0.0710 |

**Table III** $Mg_{1.2}C_{1.44}B_{0.16}Ni_3$: Lattice parameters and Unit cell volume: $a = 3.8173$ (3) Å, Vol. 55.625 Å$^3$ and $\chi^2 = 4.49$

| Atom | Site | x | y | z | Site Occupancy |
|------|------|--------|--------|--------|----------------|
| Mg   | 1a   | 0.0000 | 0.0000 | 0.0000 | 1.0277 |
| C    | 1b   | 0.5000 | 0.5000 | 0.5000 | 0.9814 |
| Ni   | 3c   | 0.0000 | 0.5000 | 0.5000 | 1.0256 |
| B    | 1b   | 0.5000 | 0.5000 | 0.5000 | 0.1013 |

**Table IV** Superconducting transition ($T_c$), lower critical field ($H_{c1}$) and upper critical field ($H_{c2}$) for pure, 5% and 10% substituted MgCNi$_3$.

| Sample Name | $T_c$ (K) | $H_{c1}$ (Oe) at 2K | $H_{c2}$ (kOe) at 2K |
|---|---|---|---|
| Mg$_{1.2}$C$_{1.6}$Ni$_3$ | 7.3 | 150 | 91.70 |
| Mg$_{1.2}$C$_{1.52}$B$_{0.08}$Ni$_3$ | 6.9 | 160 | 58.44 |
| Mg$_{1.2}$C$_{1.44}$B$_{0.016}$Ni$_3$ | 6.3 | 200 | 42.50 |

# Figure Captions

**Figure 1** Observed (semi solid circles) and calculated (solid lines) *XRD* patterns of Mg$_{1.2}$C$_{1.6-x}$BNi$_3$ (x = 0.0, 0.08 and 0.16) compounds at room temperature. Solid lines at the bottom show the difference between the observed and calculated patterns. Vertical lines at the bottom show the position of allowed Bragg peaks.

**Figure 2** DC magnetization in *ZFC* (zero-field-cooled) and *FC* (field-cooled) situation for all Mg$_{1.2}$C$_{1.6-x}$B$_x$Ni$_3$ (x = 0.0. 0.08 and 0.16) samples at 10 Oe.

**Figure 3** Isothermal magnetization (*MH*) plots at 2, 4 and 6 K in low field range of <1000 Oe for (a) Mg$_{1.2}$C$_{1.6}$Ni$_3$ (b) Mg$_{1.2}$C$_{1.52}$B$_{0.08}$Ni$_3$ and (c) Mg$_{1.2}$C$_{1.44}$B$_{0.16}$Ni$_3$, the lower critical field ($H_{c1}$) is marked.

**Figure 4** Isothermal magnetization *(MH)* for real part of AC susceptibility (*M'*) with applied field up to 140 kOe at 2 K for (a) Mg$_{1.2}$C$_{1.6}$Ni$_3$ (b) Mg$_{1.2}$C$_{1.52}$B$_{0.08}$Ni$_3$ and (c) Mg$_{1.2}$C$_{1.44}$B$_{0.16}$Ni$_3$, the upper critical field *($H_{c2}$)* is marked.

**Figure 1**

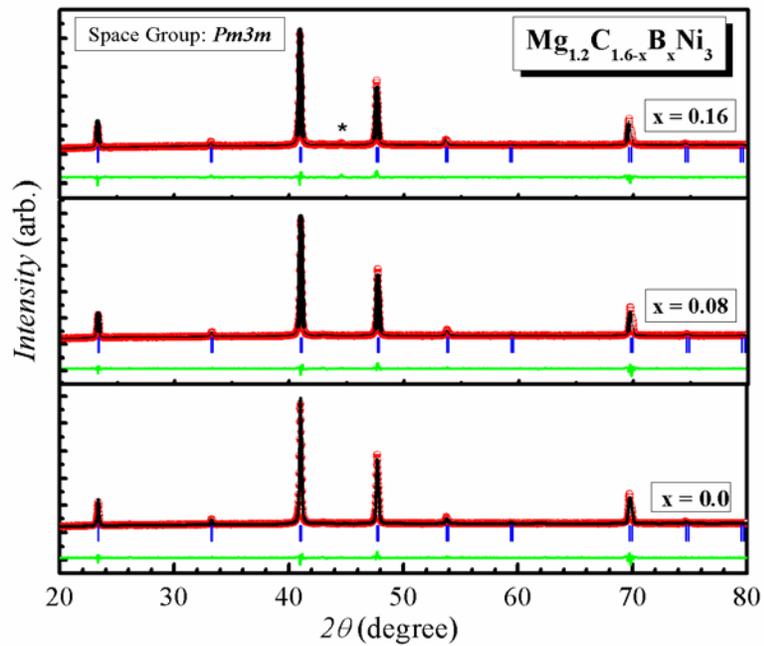

**Figure 2**

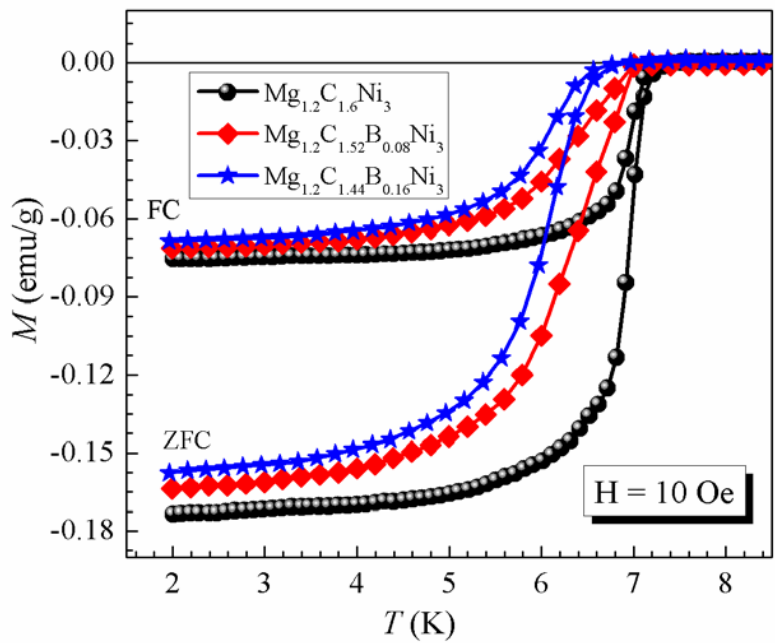

**Figure 3**

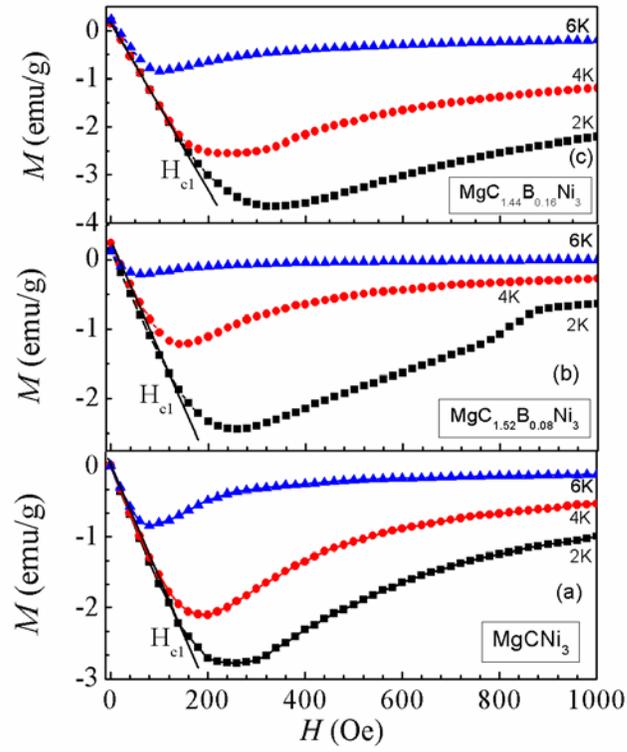

**Figure 4**

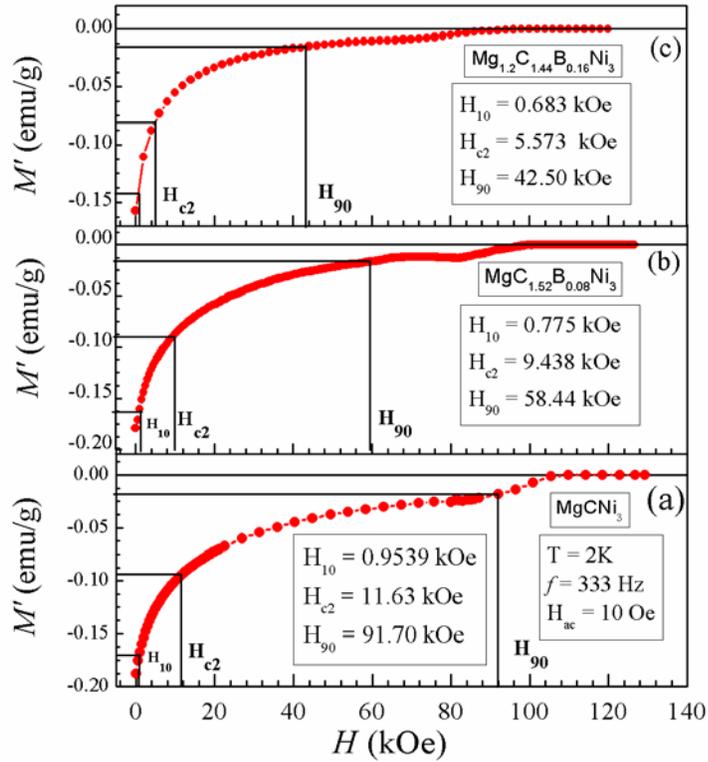